# Single-Exciton Gain and Stimulated Emission Across the Infrared Optical Telecom Band from Robust Heavily-doped PbS Colloidal Quantum Dots

Sotirios Christodoulou[1,†], Iñigo Ramiro[1,†], Andreas Othonos[2], Alberto Figueroba[1], Mariona Dalmases[1], Onur Özdemir[1], Santanu Pradhan[1], Grigorios Itskos[3], Gerasimos Konstantatos*[1,4]

1. ICFO-Institut de Ciencies Fotoniques, The Barcelona Institute of Science and Technology, 08860 Castelldefels (Barcelona), Spain
2. Laboratory of Ultrafast Science, Department of Physics, University of Cyprus, Nicosia 1678, Cyprus
3. Experimental Condensed Matter Physics Laboratory, Department of Physics, University of Cyprus, Nicosia 1678, Cyprus
4. ICREA – Institució Catalana de Recerça i Estudis Avançats, Lluis Companys 23, 08010 Barcelona, Spain

† these authors contributed equally in the work

**Materials with optical gain in the infrared are of paramount importance for optical communications, medical diagnostics[1] and silicon photonics[2,3] . The current technology is based either on costly III-V semiconductors that are not monolithic to silicon CMOS technology or Er-doped fiber technology that does not make use of the full fiber transparency window. Colloidal quantum dots (CQD) offer a unique opportunity as an optical gain medium[4] in view of their tunable bandgap, solution processability and CMOS compatibility. Their potential for narrower linewidths[5] and the lower-than-bulk degeneracy[6] has led to dramatic progress towards successful demonstration of optical gain[4], stimulated emission[7] and lasing[8,9,10] in the visible part of spectrum utilizing CdSe-based CQDs. Infrared Pb-chalcogenide colloidal quantum dots however exhibit higher state degeneracy and as a result the demonstration of optical gain has imposed very high thresholds.[11,12] Here we demonstrate room-temperature, infrared stimulated emission, tunable across the optical communication band, based on robust electronically doped PbS CQDs, that reach gain threshold at the single exciton regime, representing a four-fold**

**reduction from the theoretical limit of an eight-fold degenerate system and two orders of magnitude lower than prior reports.**

Low threshold, band-edge amplified spontaneous emission (ASE) in CQDs has been at the center of intensive research over recent years as a prerequisite towards the demonstration of CQD lasing.[10,13] Engineered CQDs with suppressed Auger[7] and photodoping[14] have allowed the realization of low threshold ASE at the single exciton regime, in the visible, for CdSe CQD systems that possess a two-fold degeneracy value. Nevertheless, low-threshold band-edge ASE in the near-infrared (NIR), based on colloidal quantum dots, has remained a challenge due to the high degeneracy of Pb-chalcogenide CQDs. PbS CQDs is the material of choice for solution processed infrared optoelectronics with successful demonstrations in LEDs,[15] solar cells[16,17] and photodetectors[18,19]. The 8-fold degeneracy of PbS(e) CQDs, however, has hindered the demonstration of low-threshold optical gain and lasing, at room temperature, in the infrared across the telecommunications wavelength bands.[11, 12]

We posited that a CQD film, robustly n-doped in the heavy doping regime, can address this challenge by utilizing the doping electrons present in the first excited state of the CQDs (conduction band) to reach the population inversion condition at reduced pumping fluence. To test our hypothesis, we developed a robust electronic n-doping method for PbS CQDs in the heavy doping regime. The doping mechanism takes place by Iodine substitution of surface sulphur sites on (100) exposed surface facets (Figure 1a). DFT calculations predict n-type doping upon iodide substitution on PbS (100) surfaces, as illustrated in Figure 1b. Iodide binding on (111) surfaces on the other hand serves as passivant without causing any strong doping effects[20] (Supplementary Section 1).

Bleach of the first exciton transition as a result of Pauli blocking is one of the main signatures of successful population of the CB.[21,22] However, bare iodide-exchanged samples did not show any absorption bleaching (see Supplementary Section 2). The absence of heavy doping, evidenced by the lack of bleaching in the absorption measurements is ascribed to the presence of oxygen and water in the film upon exposure to ambient conditions. Oxygen and water have been reported as effective p-type dopants in lead chalcogenides.[23,24] In order to preserve heavy doping in our films under ambient conditions, we submitted our samples to atomic layer deposition (ALD) of alumina with the purpose of not only preventing oxygen to further incorporate in the film but also to impede undesired p-type doping effects of the oxygen/water adsorbates, pre-existent in the films upon their formation. Infilling with alumina (supported by XPS results in Supplementary Section 3), hence, is crucial to preserve robust n-type doping provided by iodide ligand-exchange procedure. Control samples of thiol-based ligand exchange chemistry did not show any signatures of heavy doping upon ALD encapsulation, further corroborating the role of iodine as the n-type dopant (see Supplementary section 2).

Figure 1d shows exciton peak bleach after alumina deposition, indicating that the ALD process was successful in maintaining heavy doping at ambient conditions. By analysing the bleach in the absorption, we have been able to quantify the doping (number of electrons in the CB per dot, see Methods) in our samples, as shown in Figure 1e. Small dots have an octahedral shape with Pb-rich (111) facets, while, as the dot diameter increases, their morphology evolves progressively to a cuboctahedron that has six sulphur-rich (100) facets.[25] Because (100) surfaces are progressively exposed with increasing PbS CQD size, the doping efficacy of the process increases with the size of the dots. Figure 1e shows that particles smaller than 4 nm in diameter do not undergo this doping process due to the lack of (100) exposed facets, whereas in particles with an exciton peak of more than 1800 nm (~7.5 nm in diameter) the conduction band is fully filled with 8 electrons (Figure 1e) in accordance with the eight-fold degeneracy of PbS CQDs. This is also evidenced by ultraviolet photoelectron spectroscopy (UPS) (see Supplementary section 4) according to which the Fermi level of small high-bandgap CQDs upon doping does not reach close to the conduction band (CB) in contrast to the large CQDs where the Fermi level crosses the CB, suggestive of heavily doped material. It is noteworthy that the doped CQD films are stable at room temperature and ambient conditions for a period of more than 2 months (see Supplementary Section 5).

To shed further evidence on the role of iodine substitution in the doping process we have performed XPS measurements of PbS CDQ films, both with the original oleate ligands and with exchanged iodide ligands (see Supplementary Section S6). Quantitative analysis of the lead and sulphur data (see Supplementary Table S1) show that the Pb/S ratio increases after ligand exchange, consistent with substitution of sulphur by iodine. Moreover, as the particle size increases (wherefore more sulphur atoms are available at the surface) the relative increase in the Pb/S ratio after ligand exchange is more pronounced. In contrast, this doping mechanism -via iodide substitution- becomes ineffective in small PbS CQDs whose surface comprises Pb-rich (111) exposed facets [26,27].

To verify our hypothesis of reaching single exciton gain threshold in doped PbS CQD films we performed transient absorption (TA) studies in undoped PbS CQD films as well as a series of doped PbS CQD films with variable initial electron occupancy doping $<N>_D$, determined by their size (Figure 1e). Henceforth, we have employed a hybrid ligand exchange treatment based on $ZnI_2$/MPA that caters for equally effective doping (as shown in Figure 1e) as well as serves better passivation of the CQDs and higher photoluminescence, as previously reported.[15] The undoped PbS CQD films demonstrate optical gain threshold $<N>_{thr}$ -expressed in excitons per dot of four- as expected from the 8-fold degeneracy (Figure 2a). Upon doping, the $<N>_{thr}$ reduces and for the case of initial doping $<N>_D$ of 5.4 the $<N>_{thr}$ reaches a value of 0.9 excitons per dot (Figure 2b). This four-fold reduction of the gain threshold upon doping outperforms the two-fold reduction reported in CdSe based CQD systems[14]. By varying the initial doping of the CQD films, according to the size-doping dependence shown in Figure 1c, we have measured $<N>_{thr}$ values of 4.7, 2.3, 1 and 0.9 excitons per dot for QD sizes with diameter (initial doping) of: 5.0 nm ($<N>_D$ = 1.4), 5.6 nm ($<N>_D$ = 3.4), 5.9 nm ($<N>_D$ = 4.4) and 6.2 nm ($<N>_D$ = 5.4) respectively (Figure 2c). This finding further corroborates our hypothesis of the effect of doping on the optical gain threshold.

The corresponding transient absorption of the undoped and doped films (Figure 2d-e) shows a mono-exponential decay for $<N>$ below the gain threshold while above the gain threshold the lifetime traces are fitted with a bi-exponential decay. The nearly mono-exponential lifetime below the gain threshold is ranging for both doped and undoped samples from 300 - 600 ps which we tentatively attribute to band-edge recombination. On the other hand, at the optical gain regime, a second ($\tau_2$) fast component is rising. We assign the fast $\tau_2$ component to the gain lifetime while the long $\tau_1$ component is likely due to band-edge recombination (100 – 300 ps) (see Supplementary Section 6). Furthermore, extracting the gain lifetime from the transient absorption data at the probing wavelength of the highest gain value (Figure 2f) we found an average gain lifetime of 27 ps in doped CQD films while the error bar represent the lowest and the highest values obtained in different pumping photon densities (see Supplementary Section 6). Slightly longer average gain lifetime of 35 ps is observed in undoped samples. It is noteworthy, that despite the presence of doping, Auger processes do not prevent reaching the gain regime. This can be due to the fact that gain lifetime is faster than the Auger and therefore competes favourably to it and/or Auger process is suppressed in this system. To shed further insights, having conductive films, we have performed transient photoconductivity measurements[28] (see Supplementary Section 7) that yield a very low value for the Auger coefficient of $10^{-31}$ $cm^6 s^{-1}$, lower than prior reports for PbS CQDs .[29] This is likely due to the conductive nature of those films.[28]

Optical gain is a prerequisite for stimulated emission. Having achieved this, next we performed amplified spontaneous emission (ASE) measurements of our thin films. In line with the TA measurements we observed stimulating emission from both doped and undoped samples (see

Supplementary Section 8). We calculated the stimulated emission occupancy threshold $<N>_{thr}$ by fitting the integrated PL spectra (see Supplementary Section 9) from the power dependent S-curves of all the samples. In Figure 3 a,b we plot the integrated ASE peak as a function of exciton occupancy and the respective power dependence measurement of two representative PbS sizes of 5.4nm ($<N>_D = 2.7$) and 5.8nm ($<N>_D = 4.0$) (Figure c-f). All the undoped samples have an $<N>_{thr}$ of 4, in agreement with the TAS measurements of Figure 2, and the ASE signal is saturated when 8 electrons have fully populated the conduction band. Power dependence measurement of those samples are shown in Figure c-f positioning the stimulating emission peak in wavelengths above 1500 nm. The sharp ASE peak has an average FWHM of 14 meV, characteristic of a stimulated emission process and comparable with reported values in the visible from CdSe-based systems.[14] To our knowledge, this is the first report of CQD ASE in the infrared characterized by ASE saturation and spectral narrowing, essential features of ASE that had remained elusive hitherto[11,30]. The stimulated emission threshold occupancy is summarized in Figure3g. The undoped samples preserve a constant value of $<N>_{thr}$ of 4 independent of their size, whereas in the case of doped CQDs increasing their size -and thereby the initial doping occupancy- the stimulated emission threshold decreases to a minimum value of 1.3 excitons in agreement with the transient absorption measurements.

A figure of merit of paramount importance for applications in optical amplification and lasing is the net modal gain of the material. We have experimentally measured the net modal gain $g_{modal}$ using the variable stripe length (VSL) technique (see Supplementary Section 10). We report an average $g_{modal}$ of 30 cm$^{-1}$ nearly constant for all the undoped samples (Figure 3h). Upon doping, $g_{modal}$ increases up to a value of 114 cm$^{-1}$ for an $<N>_D$ of 4, when the conduction band is half filled. This modal gain value outperforms prior reports from HgTe infrared CQDs with $g_{modal}$ of 2.4 cm$^{-1}$ based on trap-to-band ASE[31], Er-doped fiber systems with values in the range of 0.01 – 0.1 cm$^{-1}$ and compares favourably to costly epitaxial III-V multi quantum well and quantum dot systems[32] . For the first time, we demonstrate nearly full coverage in the whole optical fiber communication spectrum (Figure 3i) from a solution processed material, extending beyond the spectral coverage of Er-doped fiber systems. The tunable spectral coverage across the infrared taken together with the high gain values and the low threshold represent a significant advance towards the deployment of infrared CQD solution processed lasers. Most importantly, ASE has been demonstrated, for the first time, in conductive films of CQDs, of paramount practical importance for the realization of electrically pumped CQD lasers.

**Acknowledgements**

The authors acknowledge financial support from the European Research Council (ERC) under the European Union's Horizon 2020 research and innovation programme (grant agreement no. 725165), the Spanish Ministry of Economy and Competitiveness (MINECO) and the 'Fondo Europeo de Desarrollo Regional' (FEDER) through grant TEC2017-88655-R. The authors also acknowledge financial support from Fundacio Privada Cellex, the program CERCA and from the Spanish Ministry of Economy and Competitiveness through the 'Severo Ochoa' Programme for Centres of Excellence in R&D (SEV-2015-0522). S.C. acknowledges support from a Marie Curie Standard European Fellowship (NAROBAND, H2020-MSCA-IF-2016-750600). I. Ramiro acknowledges support from the Ministerio de Economía, Industria y Competitividad of Spain via a Juan de la Cierva fellowship

**Competing interests**

G.K., I.R. and S.C. have filed patent applications related to this work.

**Author Contributions**

S.C. proposed and designed the ASE experiments, M.D. and S.C. developed the synthesis of PbS QDs, S.C. fabricated the devices, I.R. developed and optimized the doping process of the CQDs, I.R. and O.O. perform the absorption measurements and atomic layer deposition, I.R. analyzed the absorption spectra, S.C performed the XPS analysis, A.F. performed the theoretical calculations, A.O. performed the transient absorption measurements, A.O and S.C. analysed the transient absorption data, S.C. performed and analysed the ASE experiments, S.P. performed and analysed the photocurrent measurements, G.I. contributed in the discussion of the results, G.K. designed the experiments and directed the study, S.C. I. R, and G.K. wrote the manuscript with the contribution of all the authors.

**METHODS**

**PbS CQDs synthesis**

PbS QDs synthesis was adapted from a previously reported multi-injection procedure. Briefly, 0.446 g lead(II) oxide (PbO, 99.999%Pb, Strem Chemicals), 50 mL 1-Octadecene (ODE, 90%, Alfa Aesar) and 3.8 mL oleic acid (OA, 90%, Sigma Aldrich) were introduced in 3-neck, round bottom flask and degassed overnight, under vacuum at 90 °C. Then the reaction temperature was increased at 95-100°C under Argon and 60µL of Hexamethyldisilathiane ((TMS)$_2$S, Sigma Aldrich) diluted in 3ml of ODE was swiftly injected. After 6 minutes, a second solution of 75µL (TMS)$_2$S in 9ml ODE was injected dropwise in a rate of 0.75mL/min. The reaction was constantly monitored with aliquots and is was stop when at the desirable QD size. At that point both the heating and the injection was stopped and the solution was let cool down slowly at room temperature. QDs were purified three times by precipitation with anhydrous acetone and ethanol and re-dispersed in anhydrous toluene. Finally, the concentration was adjusted to 30 mg/mL and the solution was bubbled with N$_2$ in order to minimize to oxidation of the QDs.

**Doped PbS CQD films**

The ad-hoc PbS CQDs (30 mg/ml) were spin-cast onto soda-lime glass substrates (1 cm x 1 cm) at the speed rate of 2500 rpm for 20s. The film was treated with ZnI$_2$/MPA (7mg/ml of ZnI$_2$ dissolved in 0.01% of MPA in Methanol) solution for 5s and the spin-coater was started again to dry the film, while 300µL was MeOH were drop-casted to wash away the remain ligands. This procedure was repeated till the film thickness of ~110nm (4-5 layers). The film thickness was measured with profilometer. The PbS CQD films were doped after the capping with Al$_2$O$_3$ with atomic layer deposition (ALD).

**Transmission and absorption measurements**

Room-temperature absorption measurements were taken under ambient atmosphere, using a Cary 5000 UV-Vis-NIR spectrometer.

**Measurements of doping level by optical measurements**

A baseline correction was applied to the absorption measurements in order to allow proper comparison between films before and after the ALD process. Since, the $1S_e$ states of PbS are eight-fold degenerated (including spin), the number of electrons in the CB per dot, $<N>_D$, can be calculated in a straightforward manner from the bleach of the first exciton transition. If we define $I_1$ and $I_2$ as the integrated absorption strength of the excitonic transition of the undoped and doped samples, respectively, then $<N>_D = 8(1 - I_2/I_1)$. Note that by saying undoped sample, we are assuming that the doping (whether p-type or n-type) of the samples without alumina is low enough to consider full valence band and empty conduction band.

**X-ray and Ultra-Violet photoelectron spectroscopy measurements**

XPS and UPS measurements were performed with a Phoibos 150 analyzer (SPECS GmbH, Berlin, Germany) in ultra-high vacuum conditions (base pressure 5E-10mbar). XPS measurements were performed with a monochromatic Kalpha x-ray source (1486.74eV) and UPS measurements were realized with monochromatic HeI UV source (21.22eV). All the peaks have been fitted with a GL(30) line shape while the Pb4f and S2p peaks are assigned according to our previous work.[1] The quantification analysis has been performed taking under consideration the whole contribution of the lead and respectively the sulphur species corrected with the relative sensitivity factors (RSF).

**Reflection measurements of CQD films**

Reflection measurements were obtained using a PerkinElmer Lambda 950 UV/Vis/NIR spectrophotometer equipped with a Universal Reflectance Accessory module.

**Transient Absorption measurements**

Transient absorption measurements were carried out using a titanium sapphire based ultrafast amplifier centered at 800 nm and generating 45 fs pulses at a repetition rate of 1 kHz. The optical setup utilized was a typical pump-probe non-collinear configuration. The main part of the fundamental energy from the amplifier was directed into a half wave plate and a thin film polarizer system to control the energy of the excitation pulse incident of the sample. The optical path of the pump beam included an optical chopper allowing the use of phase-sensitive detection thereby improving the signal-to-noise ratio. An optical parametric amplifier pumped with approximately 1mJ of the fundamental 800nm energy was used to generate the probe beam with wavelengths ranging from 1200nm to 1700nm. The probe beam optical path included a precise motorized translation stage to control the optical delay between the pump the probe beam. The probe beam was directed on the sample within the excitation area of the pump beam where changes in transmission and reflection were recorded simultaneously using lock-in amplifiers.

**Amplified Spontaneous Emission measurements**

For the ASE measurements the ultrafast laser pulse at 800nm was directed through a cylindrical lens (focus length 15 cm) onto the sample at normal incidence. The stripe width was 700 µm while the stripe length was measured for every measurement in order to calculate the occupancy values (average value of 0.35 cm ± 0.05). The thickness the PbS CQD films was ~110nm in order to avoid over-estimation of the occupancy per dot. The emission was collected perpendicular to the incident beam using 6 cm focusing lens (5 cm diameter) and coupled into an Andor spectrometer (Shamrock SR-303) equipped with an InGaAs camera (iDus).

**Photocurrent measurements**

The QD thin films were prepared on the top of the Si/SiO$_2$ substrate with patterned Au electrode following the standard EDT (0.2%) and ZnI$_2$/MPA ligand treatments. The distance between two Au electrodes was fixed at 10 µm. 637 nm wavelength continuous laser (Vrotran stradus 637) was used to excite the QD films. All the measurements were performed in ambient conditions using an Agilent B1500A semiconducting device analyser.

**Computational details**

Density functional calculations of PbS have been performed by periodic plane-wave code Vienna ab initio simulation package VASP[2]. All structures have been optimized using the Perdew-Burke-Ernzerhof (PBE)[3] exchange-correlation functional, one of the most widely employed functionals of the generalized gradient approximation (GGA) family. It is important to mention that pure GGA functionals tend to underestimate electronic properties of materials such as band gaps. In order to account for the best possible and detailed description of the electronic structure of PbS, single-point calculations using the Heyd-Scuseria-Ernzerhof (HSE06)[4] hybrid exchange-correlation functional containing a fraction of nonlocal Fock exchange has been applied on the preoptimized PBE geometries. A plane-wave basis set with a 315 eV cutoff for the kinetic energy and a projector-augmented wave description of core−valence electron interactions were employed.[5] The one-electron Kohn−Sham states were smeared by 0.1 eV using Gaussian smearing. Finally, converged energies were extrapolated to zero smearing. All calculations were performed using a k-point Monkhorst-Pack[6] mesh of 3×3×1 in the reciprocal space for the unit cell of PbS. Relaxation of all atoms in the calculated models was carried out during the geometry optimization until forces acting on each atom became less than 0.01 eV/Å. In addition, the electron density was converged using a threshold of 10-6 eV for the total energy. No corrections for the zero-point energies were applied.

Two slab models consisting on 2×2×1 supercells were chosen to study PbS, one for the (100) surface and another for the (111) surface. The model of the stoichiometric (100) surface contains 32 atoms arranged in 4 layers, each layer formed by combination of Pb and S atoms, resulting in a nonpolar surface. Meanwhile, the (111) surface has been modelled using 28 atoms arranged in 3 Pb-S bilayers and an extra layer of Pb atoms, giving rise to two Pb terminations. We considered the PBE optimized lattice parameter to model all slabs (6.004 Å)[7], which is slightly larger than the reported experimental one of 5.929 Å.[8]

The interaction of iodine with PbS surfaces has been modelled in two different ways, doping and adsorption of I atoms, respectively. Doping of PbS by iodine was modelled by substituting one of the S atoms located on the outmost layer of the (100) surface by an iodine. In turn, adsorption of iodine on the (111) surface has been modelled by covering the two Pb termination with I atoms, leading to a 100% coverage situation. Only the hexagonal-close-pack site has been considered in the present study. In order to account for the possible electron transfers emerging due to these two different processes, all calculations were spin-polarized.

FIGURES

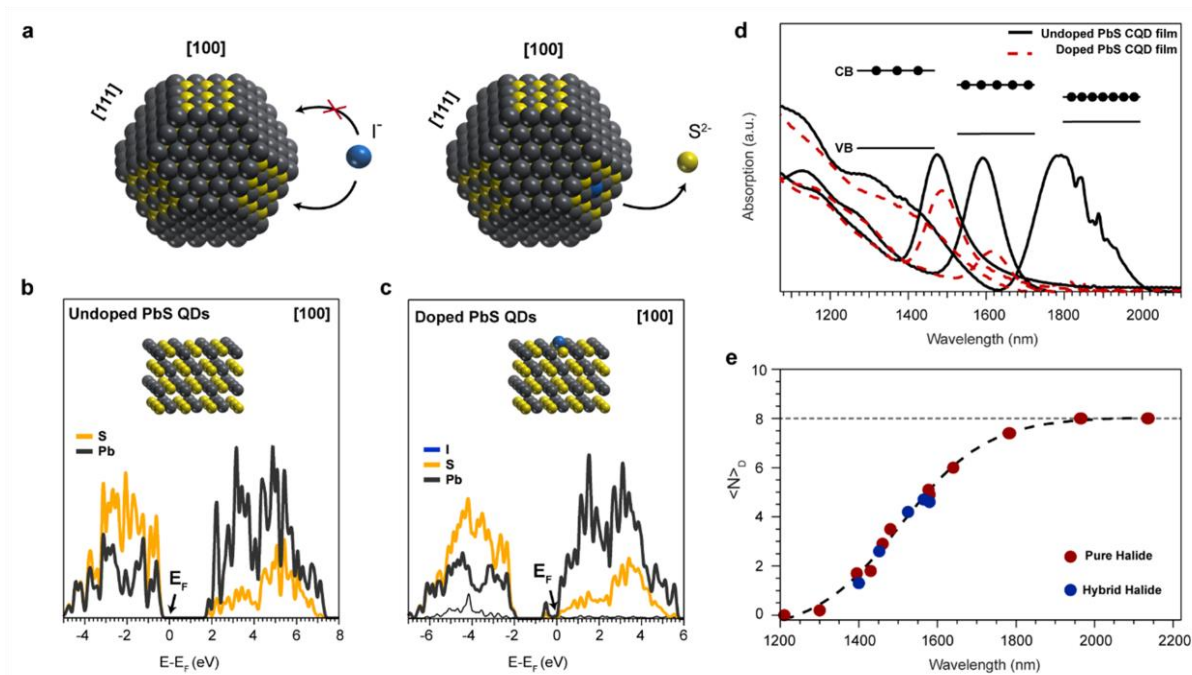

**Figure 1 | Doping in PbS CQDs films. a,** Schematic representation of the $S^{2-}$ substitution to $I^-$ in (100) surface in large, cuboctaehedral-shaped PbS CQDs **b**, Calculated density of states (DOS) of the (100) surface before and after $I^-$ substitution showing that the Fermi level, $E_F$ is shifted to the conduction band **c**, Absorption spectra of two representative PbS CQD films, before and after doping. The CQD sizes are 5.5 nm and 6.1 nm with respective exciton peaks at 1480 nm and 1580 nm (solid black lines). After doping of the CQD films the absorption bleaches (red dash lines) **e**, The number of electrons in the conduction band, upon doping, depends on the size of the CQD due to the degree of (100) surface presence that enables doping. Red dots represent the experimentally extracted number of electrons from measuring the bleaching of the absorption of the films at the exciton peak with the use of 1-ethyl-3-methylimidazolium iodide (EMII) for ligand exchange while blue dots represent the $ZnI_2$/MPA hybrid ligand treated films that are used for ASE and gain measurements. Both ligand treatments are equally effective in the doping of the CQDs. The data have been fitted with a sigmoidal function (black dash line) as a guide to the eye.

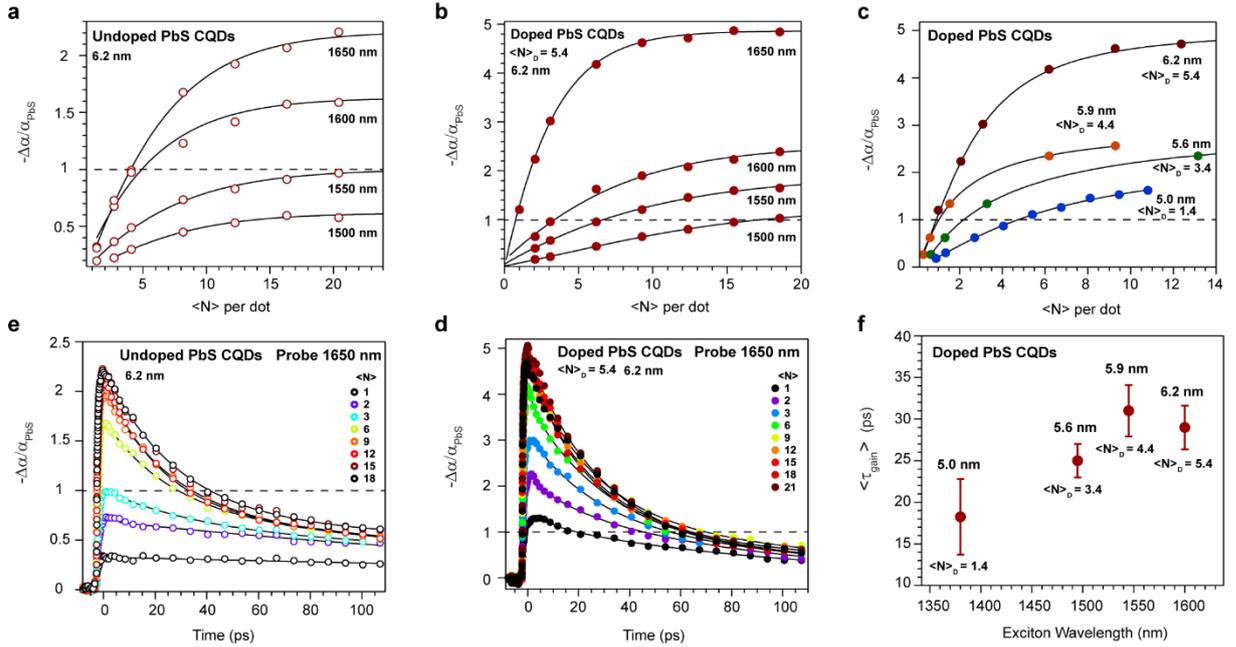

**Figure 2 | Optical gain and transient absorption in PbS CQD solids.** The gain spectra have been obtained in various probing wavelengths and different pump fluencies which are presented as exciton occupancy, <N> for both doped and undoped CQD films. The optical gain arises at the point that $-\Delta\alpha/\alpha_{PbS} > 1$ (horizontal black dash line) **a**, The gain spectra of the 6.2 nm PbS CQD (exciton peak at 1600nm) shows gain threshold at $<N>_{thr} = 4$ **b**, The corresponding doped film exhibits a drastically reduced gain threshold at $<N>_{thr} = 0.9$ at 1650 nm **c,d**, The transient absorption curves of the aforementioned doped and undoped samples at the probing wavelength of 1650 nm show that in the regime below optical gain the carrier dynamics follow a mono-exponential decay correlated with the band-edge relaxation. In the optical gain regime another fast component appears which is assigned to the gain lifetime **e**, Comparison of gain spectra of various sized doped PbS CQDs films, possessing different initial doping values. The $<N>_{thr}$ reduces upon increasing initial doping. **f**, Average gain lifetime $<\tau_{gain}>$ of different CQD sizes calculated from a range of pumping intensities while the error bars present the lowest and the highest measured values.

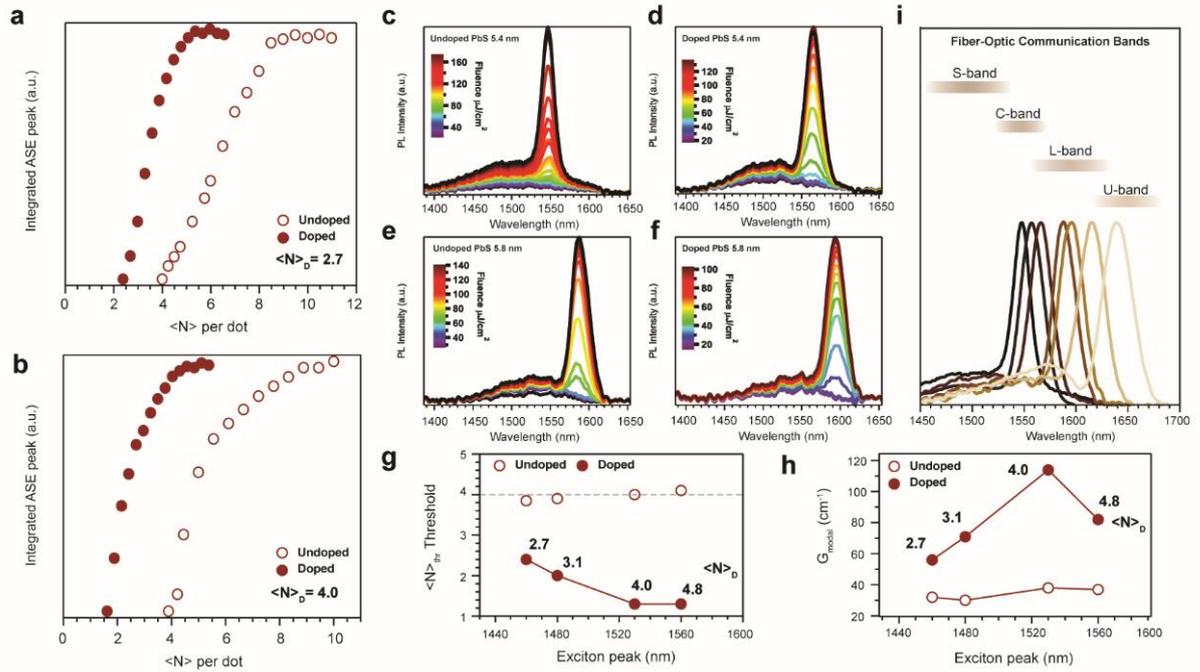

**Figure 3 | ASE and VSL measurements in PbS CQD films** We collected the ASE and the VSL spectra in a series of PbS CQD films both doped and undoped, while here we show selective data for two samples with $<N>_D$ of 2.7 and 4. We have excited our samples with a stripe-shaped laser beam at 800nm with a Ti:Saphire femtosecond laser with pulse width of 80fs and repetition rate of 1KHz at ambient conditions. **a,b**, The ASE area has been calculated by fitting the ASE peak with a Gaussian function while the photon density it is shown in terms of exciton occupancy. The solid red dot represent the integrated ASE area of doped and the hollow red dots the undoped samples. We further confirm that the undoped samples show ASE threshold $<N>_{thr} = 4$, while the doped samples show $<N>_{thr} < 4$ while both ASE signal are saturated for a total occupancy of 8 **c-f**, The corresponding PL spectra of these sample in a range of pumping intensities with a threshold of 70 µJ/cm² for the undoped and down to 25 µJ/cm² for the doped films **g**, Summarizing ASE thresholds in terms of exciton occupancy in PbS CQD of different sizes **h**, The net modal gain in the undoped films remains stable at 30 cm⁻¹ while in doped samples is increased with a peak value of 114 cm⁻¹ **i**, Collective ASE spectra from a series of PbS CQD films shows the tunability of the ASE peak from 1530 nm to 1650 nm, within the range of the fibre-optic communication bands.